\documentclass[conference]{llncs}
\pdfoutput=1

\usepackage{graphicx}
\usepackage{bmpsize}
\usepackage{xcolor}
\usepackage[export]{adjustbox}
\usepackage{cite}
\usepackage{amsmath,amssymb,amsfonts}
\usepackage{algorithmic}
\usepackage{textcomp}
\usepackage[title,header]{appendix}
\usepackage{stmaryrd}
\usepackage{tikz}
\usetikzlibrary{positioning,calc,shapes,decorations.pathreplacing,fit}
\usepackage{multirow}
\usepackage{xspace}
\usepackage{xargs}                      
\usepackage{pstricks}
\usepackage{paralist} 
\usepackage{pifont} 
\usepackage{wrapfig}
\usepackage{hyperref}
\usepackage{listings}
\usepackage{rwthcolors}
\usepackage{array}
\usepackage{booktabs}
\usepackage{subfig}
\usepackage{url}
\usepackage[nomessages]{fp}
\usepackage{cleveref}
\usepackage{longtable}
\usepackage{tabularx}
\usepackage{lscape}
\usepackage{rotating}
\usepackage{array}
\usepackage{multirow}
\usepackage{calc}
\usepackage{xspace}
\usepackage{relsize}
\usepackage{float}

\usepackage{caption}

\newcommand{\cmark}{\ding{51}}
\newcommand{\xmark}{\ding{55}}

\definecolor{palette-green}{RGB}{142,193,39}
\definecolor{palette-red}{RGB}{212,18,67}
\definecolor{palette-violet}{RGB}{162,0,255}
\definecolor{palette-blue}{RGB}{35, 102, 0}

\definecolor{codegreen}{rgb}{0,0.6,0}
\definecolor{codegray}{rgb}{0.5,0.5,0.5}
\definecolor{codepurple}{rgb}{0.58,0,0.82}
\definecolor{backcolour}{rgb}{0.95,0.95,0.92}
\definecolor{gray}{rgb}{0.4,0.4,0.4}
\definecolor{darkblue}{rgb}{0.0,0.0,0.6}
\definecolor{cyan}{rgb}{0.0,0.6,0.6}
\definecolor{codered}{rgb}{0.86,0.31,0.20}
\definecolor{whiteorange}{RGB}{255,224,171}

\lstdefinestyle{pseudocode}{
	basicstyle=\scriptsize\ttfamily,  
	commentstyle=\color{codegreen},
	keywordstyle=\bfseries,
	numberstyle=\tiny\color{codegray},
	stringstyle=\color{codepurple},
	breakatwhitespace=false,         
	breaklines=true,
	captionpos=b,                    
	keepspaces=true,
	numbers=none,
	numbersep=5pt,                  
	showspaces=false,                
	showstringspaces=false,
	showtabs=false,                  
	tabsize=2
}

\newcommand{\highlight}[1]{\textcolor{blue}{\emph{#1}}}

\newcommandx{\unsure}[2][1=]{\todo[linecolor=red,backgroundcolor=red!25,bordercolor=red,#1]{#2}}

\newcommand{\kIndTool}{CBMC+k\xspace}

\makeatletter
\renewcommand{\paragraph}{%
	\@startsection{paragraph}{4}%
	{\z@ }{-5\p@ \@plus -4\p@ \@minus -4\p@ }{-0.5em \@plus -0.22em
		\@minus -0.1em}%
	{\normalfont\normalsize\itshape}%
}
\makeatother

\newsavebox{\mybox}

\newcommand{\notes}[1]{}

\newcommand{\aref}[1]{\hyperref[#1]{Appendix}} 

\def\BibTeX{{\rm B\kern-.05em{\sc i\kern-.025em b}\kern-.08em
    T\kern-.1667em\lower.7ex\hbox{E}\kern-.125emX}}

\begin{document}

\title{Benchmarking Software Model Checkers \\ on Automotive Code}

\author{
	Lukas Westhofen\inst{1} \and
	Philipp Berger\inst{2} \and
	Joost-Pieter Katoen\inst{2}
}

\institute{
	OFFIS e.V., Oldenburg, Germany\\
	\email{lukas.westhofen@offis.de}\and
	RWTH Aachen University, Aachen, Germany\\
	\email{\{berger, katoen\}@cs.rwth-aachen.de}\vspace*{1em}
}

\maketitle

\newif\ifslicing

\newif\ifanon

\newif\ifappendix
\appendixtrue 

\begin{abstract}
This paper reports on our experiences with verifying automotive C code by state-of-the-art open source software model checkers.
The embedded C code is automatically generated from Simulink open-loop controller
models. 
Its diverse features (decision logic, floating-point and pointer arithmetic, rate limiters and state-flow systems) and the extensive use of floating-point variables make verifying the code highly challenging. 
Our study reveals large discrepancies in coverage --- which is at most only 20\% of all requirements --- and tool strength compared to results from the main annual software verification competition. 
A hand-crafted, simple extension of the verifier CBMC with $k$-induction delivers results on 63\% of the requirements while the proprietary BTC EmbeddedValidator covers 80\% and obtains bounded verification results for most of the remaining requirements.
\end{abstract}

\section{Introduction}
\label{sec:intro}
\paragraph{Software Model Checking.}
Software model checking is an active field of research.
Whereas model checking algorithms initially focused on verifying models, various dedicated techniques have been developed in the last two decades to enable model checking of program code.
This includes e.g., predicate abstraction, abstract interpretation, bounded model checking, counterexample-guided abstraction refinement (CEGAR) and automata-based techniques.
Combined with the enormous advancements of SAT and SMT-techniques~\cite{Beyer-UnifyingSMTBasedVerification}, nowadays program code can be directly verified by powerful tools.
Companies like Microsoft, Facebook, Amazon, and ARM check software on a daily basis using in-house model checkers.
The enormous variety of code verification techniques and tools has initiated a number of software verification competitions such as RERS, VerifyThis, and SV-COMP.
For software model checking, the annual SV-COMP competition is most relevant.
Launched with 9 participating tools in 2012, it gained popularity over the years with more than 40 competitors in 2019~\cite{Beyer19-svcomp}. 
It runs off-line in a controlled manner, and has several categories.
Competitions like SV-COMP have established standards in input and output format, and evaluation criteria.
Software model checkers are ranked based on the verification results, earning points for correct results while being punished for wrong outcomes.
A more recent development is the usage of witnesses to validate verification results.
Results are provided in so-called quantile plots indicating the required verification time versus the cumulative score over the benchmarks.

\paragraph{Aims of this Paper.}
This paper focuses on: \highlight{how do the SV-COMP competitors perform on automotive code?} and \highlight{how do these tools compare to proprietary tools that are tailored to such code?} 
The objective of this paper is to benchmark a rich set of participating tools in SV-COMP using two case studies from a major car manufacturer taken from~\cite{Berger-Ford}. 
In contrast to the SV-COMP, where a diverse set of open-source verification tasks ranging from small academic examples over concurrent programs up to software systems are submitted by research and development groups, we focus on an industrial grade automotive code base. 
To the best of our knowledge, such an evaluation has not been made before. 
While a set of two case studies is certainly a small benchmark in comparison, the size of the two case studies (of about 1400 and 2500 lines of embedded C code respectively), its diverse features (decision logic, floating-point arithmetic, pointer dereferencing, rate limiters, bitwise operations and state-flow systems), the rich set of (179) requirements, and the availability of verification results obtained by the proprietary software model checker BTC EmbeddedValidator, make it an interesting starting point to validate and compare various open-source software model checkers on an automotive code base.

\paragraph{Approach.}
We selected 11 software model checkers from the SV-COMP 2019~\cite{Beyer19-svcomp}, based on (a) the aforementioned characteristics of the two automotive case studies, (b) the requirements that mostly are safety properties, and (c) the availability of a license that enables an academic evaluation.
In addition, we considered a simple hand-crafted extension of CBMC~\cite{DBLP:conf/tacas/KroeningT14} with $k$-induction that is tailored to the control-flow characteristics of the two benchmarks.
We conducted
\ifslicing
\else
two
\fi
 main experiments.
The first experiment runs the 12 software model checkers on the 179 requirements, 99\% of which are invariants, and \highlight{focuses on comparing the coverage of the tools (how many requirements could be verified or refuted), and their verification time}.
The second experiment \highlight{benchmarks the open-source code verifiers against the proprietary verifier BTC EmbeddedValidator~\footnote{\url{https://www.btc-es.de/en/products/btc-embeddedplatform/}}.}

\paragraph{Our Main Findings.}
The main results of this paper are:
\begin{itemize}
\item The SV-COMP competitors are able to obtain results for at most 20\% of all requirements. Various competitors covered between 0 and 5\% only.
\item A hand-crafted, simple extension of CBMC with $k$-induction covers 63\%.
\item BTC EmbeddedValidator covers 80\% and obtains bounded verification results for 85\% of the remaining requirements.
\end{itemize}
Our results show that there is a lot of untapped optimization potential for making existing open source software model checkers more appealing and applicable to automotive code. 
Suitable benchmark candidates are currently too closely guarded by industry to be really driving scientific development. 
Therefore, the message of this paper is to emphasize the need for a synchronization between the industrial and scientific software verification communities.


\section{Preliminaries}
\label{sec:preliminaries}
\subsection{The Automotive Benchmarks}

\subsubsection{Benchmark Description.}
Both case studies involve auto-generated code of two R\&D prototype Simulink models%
\ifanon
\else
\ from Ford Motor Company%
\fi%
: the next-gen \emph{Driveline State Request} (DSR) feature and the next-gen \emph{E-Clutch Control} (ECC) feature.
The DSR and ECC features implement the decision logic for opening and closing the driveline and calculating the desired clutch torque and corresponding engine control torque of the vehicle, respectively.
The case studies are described in detail in~\cite{Berger-Ford}. 
Unfortunately, because of non-disclosure agreements, we cannot make the benchmarks publicly available; instead we give a detailed characterization of the used code in the following.

\begin{wraptable}{r}{7cm}
	\vspace{-0.4cm}
	\caption{Code metrics of the benchmarks.}
	\label{tab:metrics}
	\begin{tabular}{llrr}
		\toprule
		\multicolumn{2}{l}{\textbf{Metric}} & \textbf{DSR} & \textbf{ECC} \\
		\midrule
		
		\multicolumn{2}{l}{\emph{Complexity}} & & \\
		\cmidrule(lr){1-4}
		
		&Source lines of code & 1,354 & 2,517 \\
		&Cyclomatic complexity & 213 & 268 \\
		
		\cmidrule(lr){1-4}
		
		\multicolumn{2}{l}{\emph{Global constants}}	& \emph{77}		& \emph{274}				\\
		\cmidrule(lr){1-4}
		
		&\texttt{char}								& 12				& 8						\\
		&\texttt{char[]}							& \texttt{[12,32]} 2& 0						\\
		&\texttt{float}								& 35				& 77					\\
		&\texttt{float[]}							& \texttt{[6-12]} 9	& \texttt{[2-7]} 4		\\
		&\texttt{float*}							& 1					& 1						\\
		&\texttt{void*}								& 18				& 184					\\

		\cmidrule(lr){1-4}
		\multicolumn{2}{l}{\emph{Global variables}}	& \emph{273}		& \emph{775}			\\
		\cmidrule(lr){1-4}
		
		&\texttt{char}								& 199				& 595					\\
		&\texttt{char[]}							& \texttt{[16-32]} 3& 0						\\
		&\texttt{float}								& 46			 	& 110					\\
		&\texttt{float[]}							&  \texttt{[4-10]} 25& \texttt{[2-4]} 70	\\

		\cmidrule(lr){1-4}
		\multicolumn{2}{l}{\emph{Operations}} & \emph{5232} & \emph{10096} \\
		\cmidrule(lr){1-4}
		
		&Addition/subtraction & 133 & 346 \\
		&Multiplication/division & 52 & 253 \\
		&Bit-wise operations & 65 & 191 \\
		&Pointer dereferences & 83 & 180 \\
		\bottomrule
	\end{tabular}
	\vspace{-1cm}
\end{wraptable} 

\subsubsection{Code Characteristics.}
From the Simulink models, generated by a few thousand blocks, around 1,400 and 2,500 source lines of C code were extracted for DSR and ECC. 
Both code bases have a cyclomatic complexity of over 200 program paths. The cyclomatic complexity is a common software metric indicating the number of linearly independent paths through a program's code. 
Table~\ref{tab:metrics} presents the metrics collected on both case studies.

Constants are used to account for configurability, i.e. they represent parameters of the model that can be changed for different types of applications. 
The configurable state-space consists of 77 and 274 constants, for DSR and ECC respectively. 
Most of them are of type \texttt{float}, sometimes in a fixed-length array, as indicated by the square brackets. 
Their size range is also given in square brackets. 
Additionally, both case studies contain pointers to constant data (e.g. \texttt{const void*}). 

With a couple of hundred variables, \highlight{globals are heavily employed}. 
They are used for exchanging data with other compilation units. 
Here, the \texttt{char} type is most prevalent, taking up around three quarters of the variable count. 
\texttt{float} variables make up the remaining quarter.

The number of operations in the call graph are around $5,000$ and $10,000$ for DSR and ECC. 
While \highlight{linear arithmetic} is most prominent, we also observe a large amount of \highlight{multiplication} and \highlight{division} operations, possibly on non-constant variables. 
Challenges for software verifiers rise along with the complexity of operators used. 
Pointer and floating-point arithmetic, as well as bit-wise operations impose challenges. 
These case studies employ a variety of \highlight{bit-wise operations} such as \texttt{>>}, \texttt{\&}, and \texttt{|}, mainly on 32-bit variables. 
Such operators can force the underlying solvers to model the variable bit by bit. 
A noticeable amount of \highlight{pointer dereferences}, namely 180 and 83 occurrences, is present in the programs.

\subsubsection{Requirement Characteristics.}
The requirements originate from internal and informal documents of the car manufacturer and have been formalized by hand. 
As described in \cite{Berger-Ford}, obtaining an unambiguous formal requirement specification can be a substantial task. 
All differences between the formalization in \cite{Berger-Ford} and this work in number of properties stem from different splitting of the properties. 
For the DSR case study, from $42$ functional requirements we extracted $105$ properties, consisting of $103$ invariants and two bounded-response properties.
For the ECC case study, from $74$ functional requirements we extracted $71$ invariants and three bounded-response properties.

Invariant properties are assertions that are supposed to hold for all reachable states.
Bounded-response properties request that a certain assertion holds within a given number of computational steps whenever a given, second assertion holds.



\subsection{The Software Model Checkers}

In order to analyze the performance of open-source verifiers on our specific use case of embedded automotive C code from Simulink models, we selected a suitable subset of C verifiers based on the following criteria:

\begin{enumerate}
	\item Has matured enough to compete in the SV-COMP 2019~\cite{Beyer19-svcomp} in the \highlight{ReachSafety} and \highlight{SoftwareSystems} category.
	\item Has a license that allows an academic evaluation.
\end{enumerate}
Based on these criteria, we selected the verifiers: 2LS, CBMC, CPAChecker, DepthK, ESBMC, PeSCo, SMACK, Symbiotic, Ulti\-mate\-Auto\-mizer, UltimateKojak, and UltimateTaipan. 
The study was conducted in March 2019. We used the latest stable versions of each tool to that date. 
We also included \kIndTool (described in \Cref{subsec:kIndTool}), a variant of CBMC that enables $k$-induction as a proof generation technique on top of CBMC. 
Let us briefly introduce the selected open-source verifiers.

\paragraph{CBMC 5.11 \cite{Clarke-CBMC-2004}.} The \emph{C Bounded Model Checker} is a matured bounded model checker for C programs. CBMC takes a pre-specified bound up to which the program loops are unrolled. The resulting transition system is encoded symbolically, and finally passed to an SAT-solver. For a given bound $k$, this formula over the program states is created in the following manner, where $I$ is the initial condition, $T$ the transition relation, $s_i$ a state and $P$ the property:

{
\relscale{0.8}
\begin{equation}
\label{eq:bmc}
\mathit{BMC}_k(s_0, \dots, s_k) = I(s_0) \wedge \left(\bigwedge_{i = 0}^{k-1} T\left(s_i, s_{i+1}\right)\right) \wedge \left(\bigvee_{i = 0}^k \neg P\left(s_i\right) \right)
\end{equation}
}

\paragraph{ESBMC 6.0.0 \cite{Gadelha-ESBMC-5.0}.} The \emph{Efficient SMT-based Bounded Model Checker} was forked off of a 2008 version of CBMC and has been replacing original framework parts ever since. 
One of its goals is to directly translate to SMT-theories instead of relying on SAT-solvers. 
It furthermore supports $k$-induction. 
Here, a generalized mathematical induction is applied to program loops, where a ``look-back'' of $k$ steps is allowed for the induction hypothesis. 
The verification task can be specified as a formula over the program states:

{
\relscale{0.8}
\begin{equation}
\label{eq:kinduction}
\mathit{IND}_k(s_0, \dots, s_k) = \left(\bigwedge_{i = 0}^{k-1} T\left(s_i, s_{i+1}\right)\right) \wedge \left(\bigwedge_{i = 0}^{k-1} P\left(s_i\right) \right) \wedge \neg P(s_k)
\end{equation}
}

\paragraph{2LS 0.7.0 \cite{Schrammel-2LS-SVComp2016}.} This is another fork of CBMC that expands from bounded model checking to a multitude of verification approaches. It interprets program analysis as a problem of solving a second-order logic instance. This leads to a variety of concepts that \textsc{2ls} can employ, including (incremental) bounded model checking, $k$-induction, $k$-induction $k$-invariants, and abstract interpretation.

\paragraph{CPAChecker 1.8.0 \cite{Beyer-CPAChecker-First}.} The \emph{Configurable Program Analysis Checker} provides a framework for implementing a rich set of analysis and verification techniques. By employing an abstract analysis algorithm, it implements concrete approaches such as predicate abstraction \cite{Beyer-CPAChecker-PredicateAbstractionWithABE}, value analysis \cite{Beyer-CPAChecker-ExplicitAnalysis}, and $k$-induction \cite{Beyer-CPAChecker-kInduction}.

\paragraph{PeSCo 1.7 \cite{richter2019pesco}.} PeSCo is a recent fork of CPAChecker which exploits \emph{machine learning} to effectively select a fitting configuration for the given verification task. 

\paragraph{DepthK 3.1 \cite{Rocha-DepthK}.} DepthK uses \emph{$k$-induction on top of ESBMC} combined with an \emph{invariant-strengthening} approach.
It supports the iterative proof process by inferring possibly over-approximating invariants over polyhedral constraints.

\paragraph{SMACK 1.9.3 \cite{Rakamaric-SMACK-2014}.} Rather than being a verifier by itself, SMACK translates from the LLVM intermediate representation (IR) into the \emph{Boogie}~\cite{BarnettCDJL05} intermediate verification language (IVL). 
Corral, the default verification back end, employs bounded model checking with a goal-directed search algorithm.

\paragraph{Symbiotic 6.0.3 \cite{Chalupa-Symbiotic}.} Symbiotic applies program instrumentation, \emph{static slicing} and symbolic execution to identify counterexamples. Internally, it uses a patched KLEE version for symbolic execution and witness generation.

\paragraph{UltimateAutomizer 91b1670e \cite{DBLP:conf/cav/HeizmannHP13}.} This tool implements a trace-abstraction based on \emph{automata} in a CEGAR fashion. Its development is based on the Ultimate framework which provides access to program representation, code transformations, and SMT-solvers. It applies a CEGAR scheme until an error automaton with sufficient abstraction is found.

\paragraph{UltimateKojak 91b1670e \cite{Ermis-UltimateKojak}.} As part of the Ultimate tool chain, UltimateKojak uses CEGAR with \emph{interpolation} over multiple program paths.

\paragraph{UltimateTaipan 91b1670e \cite{Greitschus-UltimateTaipan}.} Similar to UltimateAutomizer, UltimateTaipan employs automata-based trace abstraction and CEGAR. It uses a fixed-point iteration to refine error paths until a sufficient precision is reached.

\subsection{A Simple, Tailored Variant of CBMC}\label{subsec:kIndTool}

The SV-COMP verifiers are complemented by a simple, hand-crafted extension of the bounded model checker CBMC (version 5.11) with $k$-induction.
Our implementation is tailored to the two case studies, in particular to programs with one main outermost control loop. 
Our prime motivation to consider this variant is to show the effect of a simple, almost trivial, tweak of a bounded model checker. 
The main goal of $k$-induction is harvesting the power of efficient bounded model checkers such as CBMC for proof generation. 
In this way, verifiers that natively only support bug hunting but have matured over time, can be elevated. 

\renewcommand{\ttdefault}{pcr}
\begin{figure}[ht!]
\vspace*{-0.4cm}
	\begin{center}
		\begin{minipage}{0.44\textwidth}
			\begin{lstlisting}[language=C]
extern void __VERIFIER_error();

int main() {
	initialize();
	
	
	while(1) {
	
	
		step();
		if(!property())
			__VERIFIER_error();
	}
}
			\end{lstlisting}
		\end{minipage}
		\begin{minipage}{0.05\textwidth}
			\hspace{-0.5cm}
			\rotatebox{0}{\scalebox{1.8}{$\leadsto$}}
		\end{minipage}
		\begin{minipage}{0.48\textwidth}
			\begin{lstlisting}[language=C]
extern void __VERIFIER_error();
extern void __VERIFIER_assume(int);
int main() {
	initialize();
	<@\colorbox{yellow}{set\_loop\_variables\_nondet();}@>
	unsigned int <@\colorbox{yellow}{i = 0;}@>
	while(1) {
		<@\colorbox{yellow}{\_\_VERIFIER\_assume(property());}@>
		<@\colorbox{yellow}{i++;}@>
		step();
		if(<@\colorbox{yellow}{i == k}@> && !property())
			__VERIFIER_error();
	}
}
			\end{lstlisting}
		\end{minipage}
	\end{center}
	\vspace*{-0.8cm}
	\caption[$k$-induction transformation]{The transformation that is applied in the \texttt{k}-th induction step.}
	\label{fig:kinduction}
	\vspace{-0.5cm}
\end{figure}
\renewcommand{\ttdefault}{cmtt}

Our implementation \kIndTool is realized by a straightforward code transformation~\cite{Gadelha-kInductionCodeTransformation}, see Fig.~\ref{fig:kinduction}. 
It creates a new program representing the induction step such that all input variables are set non-deterministically on entering the loop. 
It then runs the back-end verifier on both the base step -- i.e. the input file -- and the induction step. 
If the base step returns a counterexample, the tool reports \emph{False}. 
In case the induction step returns no counterexample for iteration $k$ and the base case has also reached $k$, it reports \emph{True}. 
Our two case studies do not require the forward case in~\cite{Gadelha-kInductionCodeTransformation}, thus simplifying the implementation.

\kIndTool has \highlight{severe restrictions on its input code}. 
It is targeted to embedded C programs containing one (unbounded) main loop with a strictly bounded loop body. 
The property has to be checked at the very end of every loop iteration. 
Although there exist transformations from general programs to one-loop programs, we decided to skip this step as our case studies do not exhibit nested unbounded loops. 
Evidently, \kIndTool inherits the capabilities (and deficiencies) of its back-end verifier, specifically its ability to handle large state spaces.

\kIndTool should thus not be considered a generic, widely applicable extension of CBMC with $k$-induction. 
Our prime motivation to consider this variant is to show the effect of a simple, almost trivial, tweak on a bounded model checker. 
We have taken CBMC for this variant as it performed very well in identifying counterexamples, an important trait for $k$-induction. 

\ifanon
The \kIndTool implementation is made publicly available at \url{https://github.com/icst-2020-submission/cbmc-with-kInduction}.
\else
The \kIndTool implementation is made publicly available at \url{https://github.com/moves-rwth/cbmc-with-kInduction}.
\fi

\subsection{Experimental Setup}

All experiments were performed on a machine with 192 GB RAM and two Intel Xeon Platinum 8160 processors, each containing 24 cores at 2.1 GHz. 
Our benchmark script executed ten benchmarks in parallel, giving each execution four CPU cores with a memory limit of 18 GB and a CPU-time limit of two hours.
Further details can be found in the \aref{appendix}.
Every verification was followed by two witness validation runs of CPAChecker and Ultimate. 
Conforming to the regulations of the SV-COMP 2019, the time limit for a correctness witness was two hours, whereas a violation witness had to complete within 12 minutes. 
We collected the data points:
\begin{itemize}
	\item the result; either \emph{True}, \emph{False} or \emph{Unknown},
	\item why no definite answer was given, e.g. \emph{Timeout}, \emph{Memout} or \emph{Verifier bug},
	\item the used CPU-time, in seconds,
	\item the peak memory usage, in MB,
	\item if measurable, the time spent by a SAT solver, in seconds,
	\item if measurable, the reached depth in a BMC or $k$-induction setting, and
	\item the witness validation results; either \emph{Correct} (validation result = original result), \emph{Invalid} (unparseable witness) or \emph{Unknown} (resource exhaustion, or validation result $\neq$ original result).
\end{itemize}

To keep the results comparable and the competition fair, we used the default configurations that the tool maintainers chose for the SV-COMP 2019 reachability tasks. The exact settings can be found in the \aref{appendix}.
CBMC was invoked with increasing values of $k$ by a wrapper script similar to the one employed in the SV-COMP 2019. 
For the Ultimate tool chain, a bit-precise memory model was applied\footnote{by adding \texttt{Memory\ model=HoenickeLindenmann\_Original}.} to the Boogie translator configuration. 
The witness validation processes for CPAChecker and Ultimate were set up as in SV-COMP 2019 with scaled run times where necessary. Due to the aforementioned confidentiality reasons, we cannot disclose the extracted benchmark data and verifier outputs.

\newpage
\section{Comparing the Open-Source Verifiers}
\label{sec:experimentsAndResults}
\paragraph{Coverage.} 

\begin{figure}[ht!]
	\centering
	\vspace{-0.3cm}
	\scalebox{0.53}{\input{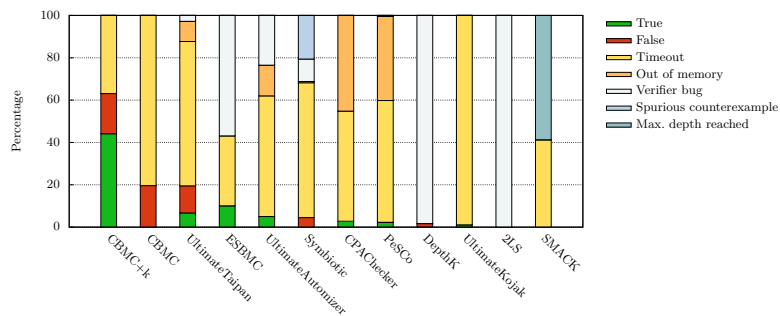}}
	\vspace*{-0.7cm}
	\caption{The overall result distribution for each software model checker, in percent.}
	\label{fig:resultdistribution}
	\vspace{-0.5cm}
\end{figure}

Fig.~\ref{fig:resultdistribution} shows the verification results of running the open-source verifiers on the two case studies%
, omitting the results of the witness validation.

%
%
\kIndTool is able to verify about 63\% of the verification tasks; CBMC and UltimateTaipan cover roughly 20\%. 
ESBMC delivers results on 10\% of the requirements.
The remaining verifiers reach a coverage of at most 5\%.
The majority of the verifiers is either able to identify counterexamples or produce proofs, but seldom both. 
2LS and SMACK cannot return a single definite result. 
The only successful witness validation was a proof of PeSCo validated by CPAChecker, indicated by \emph{True (Correct)}. 
CBMC delivered invalid witnesses on all tasks, leading it to fail the witness validation process.

Fig.~\ref{fig:resultdistribution} also indicates the reasons for \emph{Unknown} answers. 
We observe that \emph{time- and memory-outs prevail}, but a large number of verifiers exhibit \emph{erroneous behavior}. 
A detailed description of the latter issues is given in \Cref{subsec:encounteredIssues}.

\begin{figure*}
	\vspace{-0.3cm}
	\begin{minipage}{0.46\textwidth}
		\centering
		\fontsize{6pt}{6pt}\selectfont
		DSR (105)\\
		\def\svgwidth{5cm}
		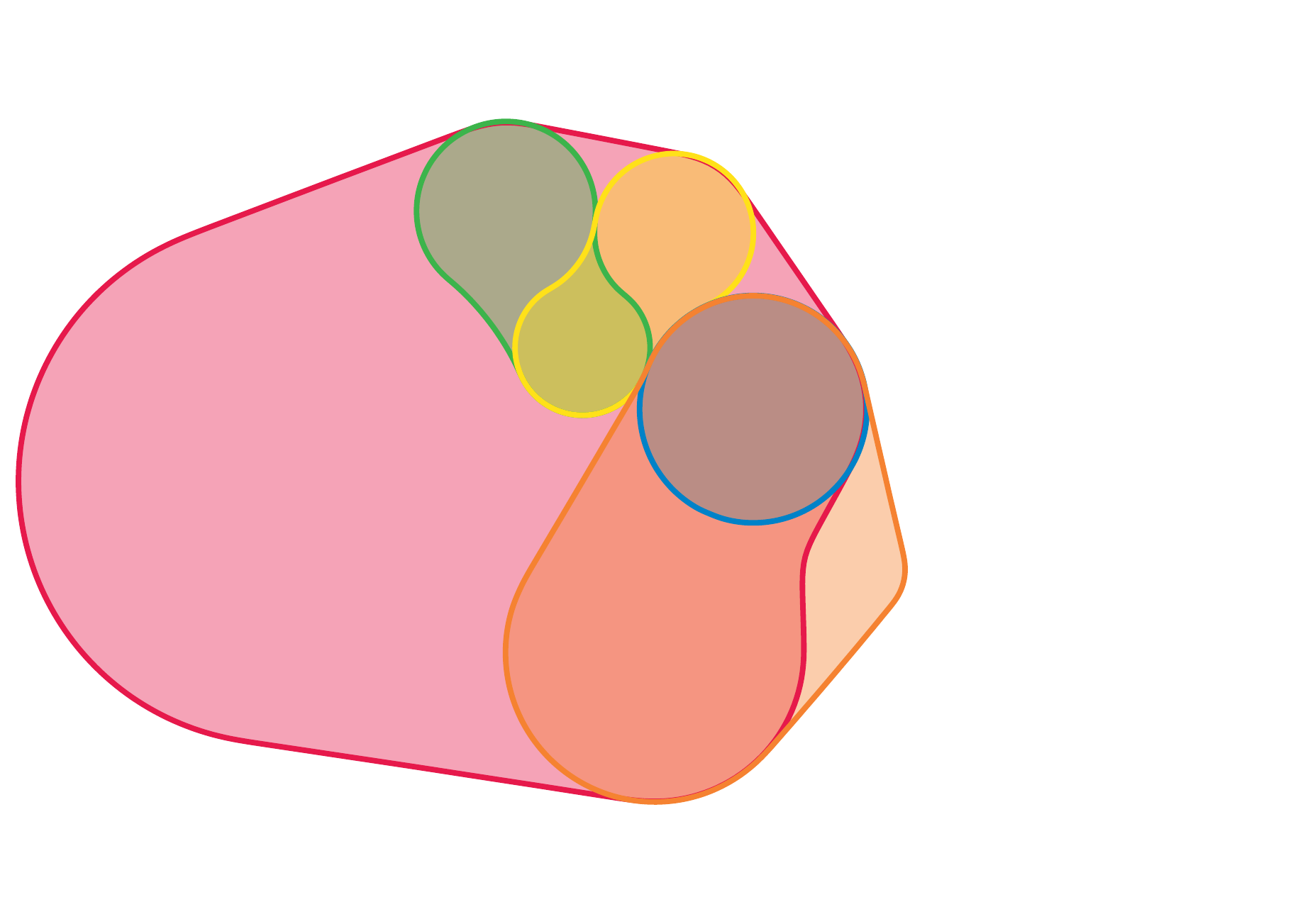
	\end{minipage}
	\begin{minipage}{0.46\textwidth}
		\centering
		\fontsize{6pt}{6pt}\selectfont
		ECC (74)\\
		\def\svgwidth{5cm}
		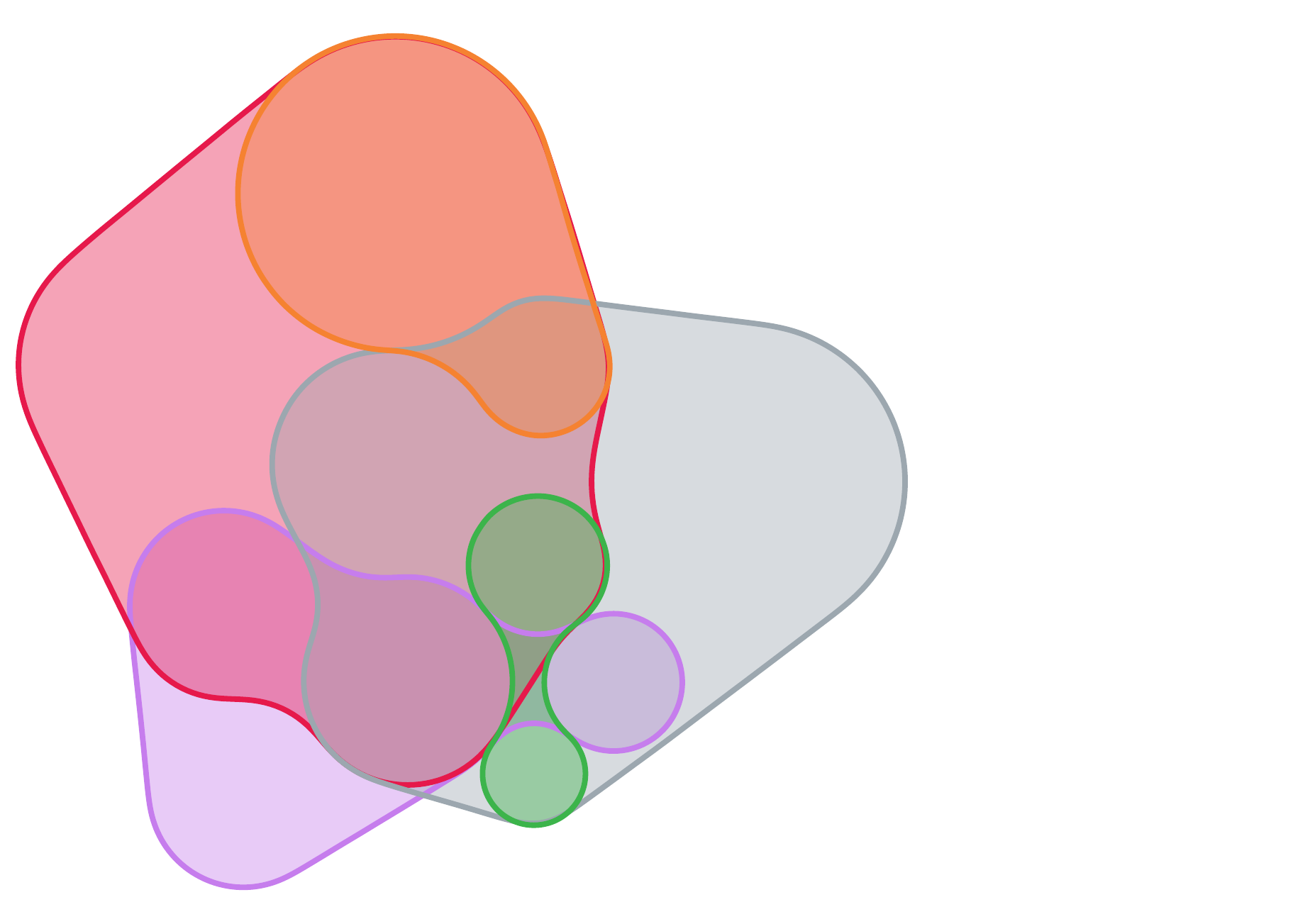
	\end{minipage}
	\vspace*{-0.3cm}
	\caption{Venn diagrams indicating the requirement coverage (i.e., a definite result was issued) by the top-five verifiers for case study DSR (left) and ECC (right).}
	\label{fig:venn}
	\vspace{-0.5cm}
\end{figure*}

To get insight into which requirements are covered by which software model checker, Fig.~\ref{fig:venn} depicts two Venn diagrams indicating the subsets of all $179$ verification tasks. 
Each area represents the set of verification tasks on which a verifier returned a definite result. 
Those areas are further divided into overlapping subareas, where a number indicates the size of this set. 
For reasons of clarity, we included only the top five verifiers for the respective case study, based on the number of definite answers.
For both case studies, there is not one verifier which covers all requirements covered by the other verifiers. 
For DSR, \kIndTool covers all but one definite results of the remaining verifiers.
In this case, CBMC was able to identify a counterexample close to the timeout.
\kIndTool exhausts its resources on this requirement as the inductive case occupies a part of the available computation time.
For ECC, UltimateTaipan, ESBMC, and \kIndTool together cover the set of all definite results. 
Note that some verifiers --- e.g. UltimateTaipan and ESBMC --- perform rather well on one case study, but lose most of their coverage on the other. In most of such cases, this is due to erroneous behavior of the verifier manifesting on just one of the two case studies.

We believe that the substantial difference in verifier coverage for the two case studies, as seen in Fig.~\ref{fig:venn}, is the result of structural differences in the benchmark code.
While the overall control-flow structure (closed loop, step-based input to output propagation) is the same for DSR and ECC, the difference in overall size and the higher number of global constants, pointers and floating-point variables make ECC imposing  different challenges.
Even a small increase in code size can lead to verifiers not even getting through costly initial preparatory steps, that, if completed, might have quickly been followed by a result.

\begin{figure}[ht!]
	\centering
	\scalebox{0.6}{\input{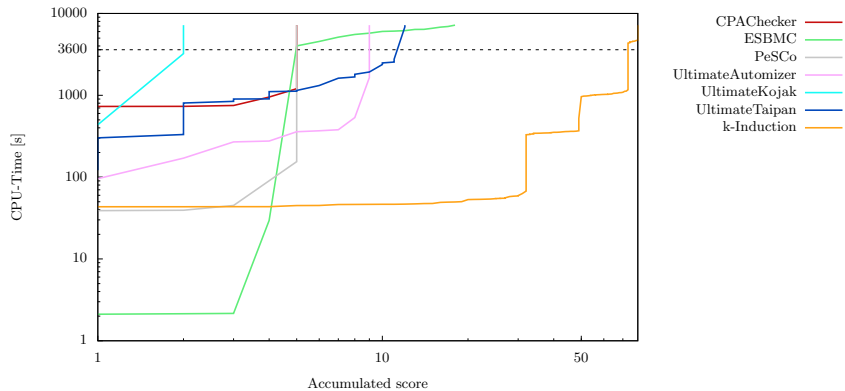}}
	\vspace*{-0.3cm}
	\caption{The quantile (log, log) plot for all verifiers except the tools 2LS, CBMC, DepthK, SMACK and Symbiotic (as they reach a zero score).}
	\label{fig:quantileplot}
	\vspace{-0.5cm}
\end{figure}

\paragraph{Quantile Plot.}
As standard in SV-COMP, a quantile plot for the results on both case studies together is depicted in Fig.~\ref{fig:quantileplot}.
Note the log-log scale.
To this end, a score is assigned to each verification run according to the SV-COMP\footnote{\url{https://sv-comp.sosy-lab.org/2019/rules.php\#scores}} scheme in Table~\ref{tab:quantilescore}. 

\begin{table}[ht!]
	\centering
	\caption{The employed scoring scheme for the quantile plots as adopted from SV-COMP.}
	\label{tab:quantilescore}
	\setlength\tabcolsep{0.3cm}
	\begin{tabular}{lcccccc}
		\toprule 
		\textbf{Verification result} & \multicolumn{3}{c}{\emph{False}} & \multicolumn{3}{c}{\emph{True}} \\ 
		\textbf{Validation result}   & \cmark & \textbf{?} & \xmark     & \cmark & \textbf{?} & \xmark    \\ 
		\midrule
		\textbf{Score}               & $+1$   & $\pm0$     & $\pm0$     & $+2$   & $+1$       & $\pm0$    \\ 
		\bottomrule
		\hline 
	\end{tabular}
	\vspace{-0.5cm}
\end{table}

The score depends on the results of the witness validation which can either be validated (verification and validation result coincide, indicated by \cmark), not validated (resource exhaustion or verification and validation result differ, \textbf{?}) or invalid (unparseable witness, \xmark). 
In absence of expected verification results, no punishments for wrong answers are given. 
In \Cref{section:compareAgainstBTC}, we compare the verification results against those obtained by the commercial verifier BTC EmbeddedValidator.


The quantile plot in Fig.~\ref{fig:quantileplot} indicates the accumulated score for all verification runs, sorted by ascending run time (x-axis), against the required CPU-time (y-axis). 
A (log, log) scale is used for improved readability.
As invalid and unvalidated counterexamples are not rewarded in this score, verifiers returning such results -- 2LS, CBMC, DepthK, SMACK and Symbiotic -- obtain a zero score.
Verifiers with a large number of proofs obtain higher scores. 
As only one witness could be validated, this aspect plays a negligible role in the scores. 
\kIndTool exhibits a higher score than other verifiers; runner-up ESBMC obtains various results only after one hour.
In general, 50\% and 90\% of the answers were given within seven and 75 minutes, respectively. 
Only few verifiers used the full time limit of two hours: The Ultimate verifiers and \kIndTool obtained many results within an hour.

\section{Benchmarking Against BTC EmbeddedValidator}\label{section:compareAgainstBTC} 
To compare the results of open-source software verifiers to a commercial tool, we additionally ran the verification tasks using BTC EmbeddedValidator (BTC for short).\footnote{Similar results were provided in~\cite{Berger-Ford}. 
We have used a more recent version of BTC EmbeddedValidator and considered $179$ rather than $112$ requirements, as requirements were split differently.}
The main purpose of this examination is the establishment of a reference point. 
This reference can subsequently be used as a foundation to interpret the applicability of the open-source verifiers to the industrial case studies.

BTC EmbeddedValidator is part of BTC EmbeddedPlatform, a commercial model-checking tool developed for industrial applications. 
It is, among others, heavily optimized for industrial embedded software --- such as the benchmarks considered in this paper --- and unsurprisingly performs very well on the ECC and DSR case studies.
This focus is also a weak point: It can not or not easily deal with memory allocation and many standard library headers usually not present in the targeted embedded code, making it unsuitable for a direct comparison on established SV-COMP benchmarks. 
Requirements can be specified directly using a pattern-based approach, see~\cite{DBLP:conf/mbmv/TeigeBH16,DBLP:conf/fmics/BergerNKAWR19}.
BTC EmbeddedValidator employs several back-end tools for verification: CBMC\footnote{A different, custom version than used in SV-COMP 2019}, iSAT3, AutoFXP, SMIBMC, and VIS.
Code transformation, static analysis, and detection of spurious, i.e., incorrect, counterexamples are done as part of the verification. 

We used BTC EmbeddedPlatform 2.3p1 under Windows 7 with 4 GB RAM and an Intel i7-6700HQ with a timeout of two hours.
While this setup is using a smaller CPU and less RAM than our experiment in Section~\ref{sec:experimentsAndResults} and is therefore incomparable, it is important to stress that we use the results of BTC only for deciding the baseline truth and do not depend on the performance (see also \ref{subsec:scoresAssumingBtc}).

\subsection{BTC EmbeddedValidator Verification Results}

Table~\ref{tab:btcresults} states the result distribution for both case studies, in percent of the 105 and 74 verification tasks, respectively. 
\begin{table}[h]
	\caption{Verification results of BTC EmbeddedValidator on both case studies, in percent of the 105 and 74 verification tasks.}
	\label{tab:btcresults}
	\setlength\tabcolsep{0.21cm}
	\resizebox{1\textwidth}{!}{%
	\begin{tabular}{lcccccc}
		\toprule 
		\textbf{Case study}          &             & DSR (105)    &                &             & ECC (74)     &                \\ 
		\textbf{Result}				 & \emph{True} & \emph{False} & \emph{Unknown} & \emph{True} & \emph{False} & \emph{Unknown} \\ 
		\midrule
		\textbf{Percentage}          & 56.2\%      & 21.9\%       & 21.9\%         & 55.4\%      & 27.0\%       & 17.6\%         \\ 
		\bottomrule
		\hline 
	\end{tabular}}
\end{table}

\begin{figure}[h]
	\centering
	\vspace{-0.3cm}
	\scalebox{0.53}{\input{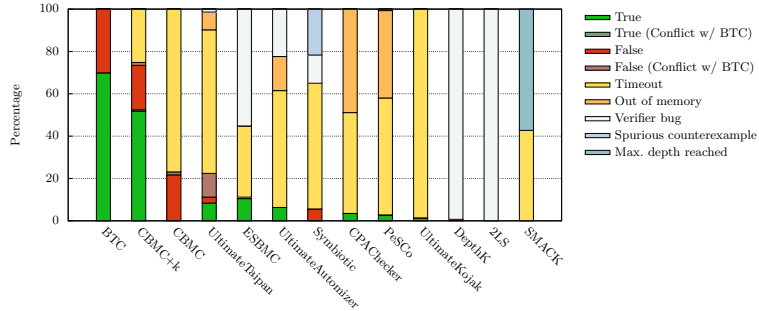}}
	\vspace*{-0.7cm}
	\caption{The verification results for each verifier, in percent of the 143 verification tasks on which BTC returned a definite result. No witness validation results are depicted, as they were previously given in Figure~\ref{fig:resultdistribution}.}
	\label{fig:resultdistribution_btc_gt}
	\vspace{-0.5cm}
\end{figure}

BTC did not return a result on 21.9\% of the DSR tasks; 91\% of which were due to reaching only bounded correctness, but no unbounded proof. 
BTC timed out on the remaining 9\%. 
Of the 17.6\% \emph{Unknown} answers for ECC, 92\% are bounded proofs, and 8\% timed out.
In comparison to the open-source verifiers, BTC takes first place in both case studies when considering the overall number of definite answers. 
As witness output is not available in BTC and wall clock times were measured, we cannot integrate BTC fairly into our scoring system, and thus refrain from calculating a quantile plot score. 
Although we were not able to determine exact CPU-times from BTC due to tool limitations, a wall clock time was collected. 
The average wall clock time of BTC on tasks where definite answers were returned amounts to $17 \pm 4$ seconds on DSR and $308 \pm 1109$ seconds on ECC.
Fig.~\ref{fig:resultdistribution_btc_gt} shows the results for all 143 verification tasks on which BTC returned a definite result.
It also indicates conflicts, i.e. different outcomes than BTC EmbeddedValidator.

\subsection{Scores Assuming Correct Results by BTC EmbeddedValidator}
\label{subsec:scoresAssumingBtc}
\begin{figure}[ht!]
	\centering
	\vspace{-0.3cm}
	\scalebox{0.6}{\input{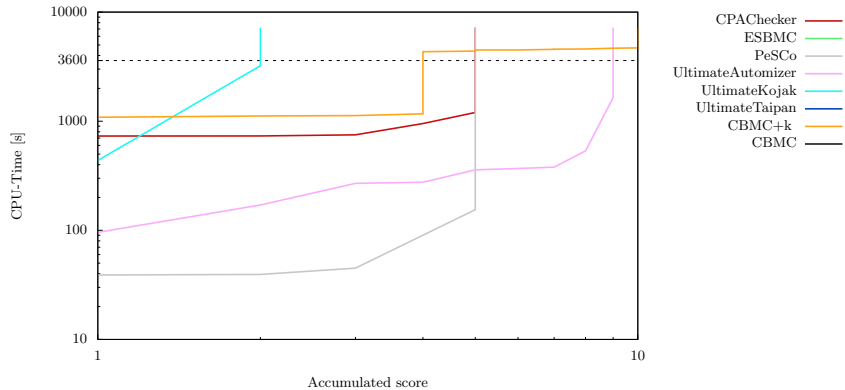}}
	\vspace*{-0.3cm}
	\caption{The quantile plot for each verifier, assuming BTC results as ground truth. The tools 2LS, DepthK, SMACK and Symbiotic have been omitted as they do not reach a	score other than zero.}
	\label{fig:quantileplot-btc-gt}
	\vspace{-0.5cm}
\end{figure}

In absence of the true verification results, let us assume the results of BTC EmbeddedValidator as a ``ground truth''. 
As this is a mature industrial tool developed over many years specifically for such industrial cases considered here, we believe that this is a reasonable assumption. 
For this, we restrict the verification tasks to those on which BTC returns a definite answer. 
We are aware of the fact that this is debatable, but given the very low number of verification results by BTC EmbeddedValidator that could be shown by other tools to be invalidated (as depicted later), this gives a quite good impression. 
We would like to point out that we are not interested in either shaming or praising specific tools, we simply are trying to provide a look at the ``big picture'' with respect to model checking certain types of industrial embedded code. 
Our assumption of using BTC EmbeddedValidator as a ground truth does certainly not imply the validity of all its results. 
But, considering the purpose of this section, it represents a sufficiently precise reference point for a comparison. 
We update the quantile plots to \emph{now punish wrong results} (i.e., results in conflict with BTC) by \emph{${-}16$ and ${-}32$ points for wrong violation and proof results}, respectively, as in the SV-COMP. 
The resulting plots are given in Fig.~\ref{fig:quantileplot-btc-gt}.

\begin{figure*}
	\vspace{-0.5cm}
	\begin{minipage}{0.46\textwidth}
		\centering
		\fontsize{6pt}{6pt}\selectfont
		DSR (82)\\
		\def\svgwidth{5cm}
		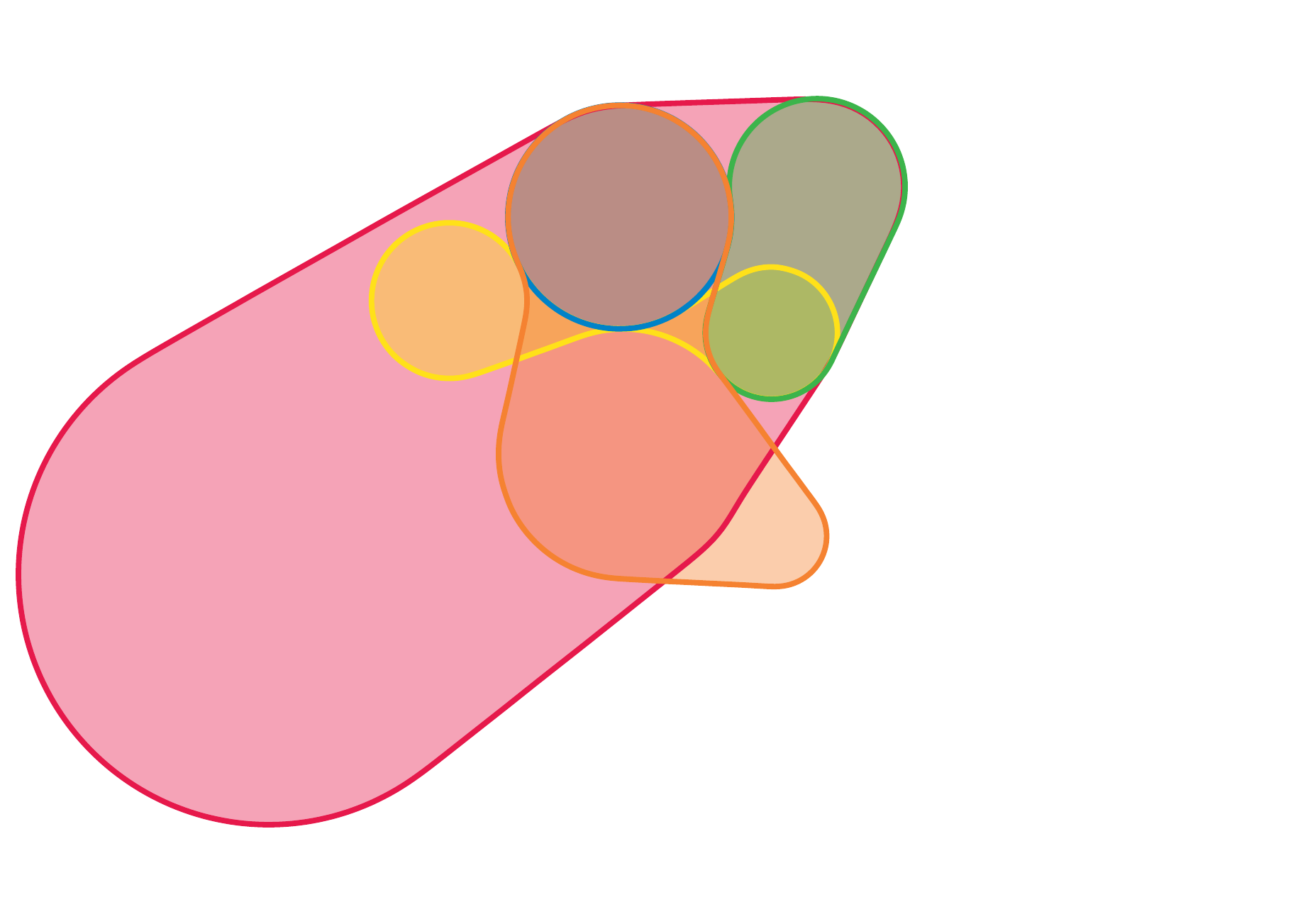
	\end{minipage}
	\begin{minipage}{0.46\textwidth}
		\centering
		\fontsize{6pt}{6pt}\selectfont
		ECC (61)\\
		\def\svgwidth{5cm}
		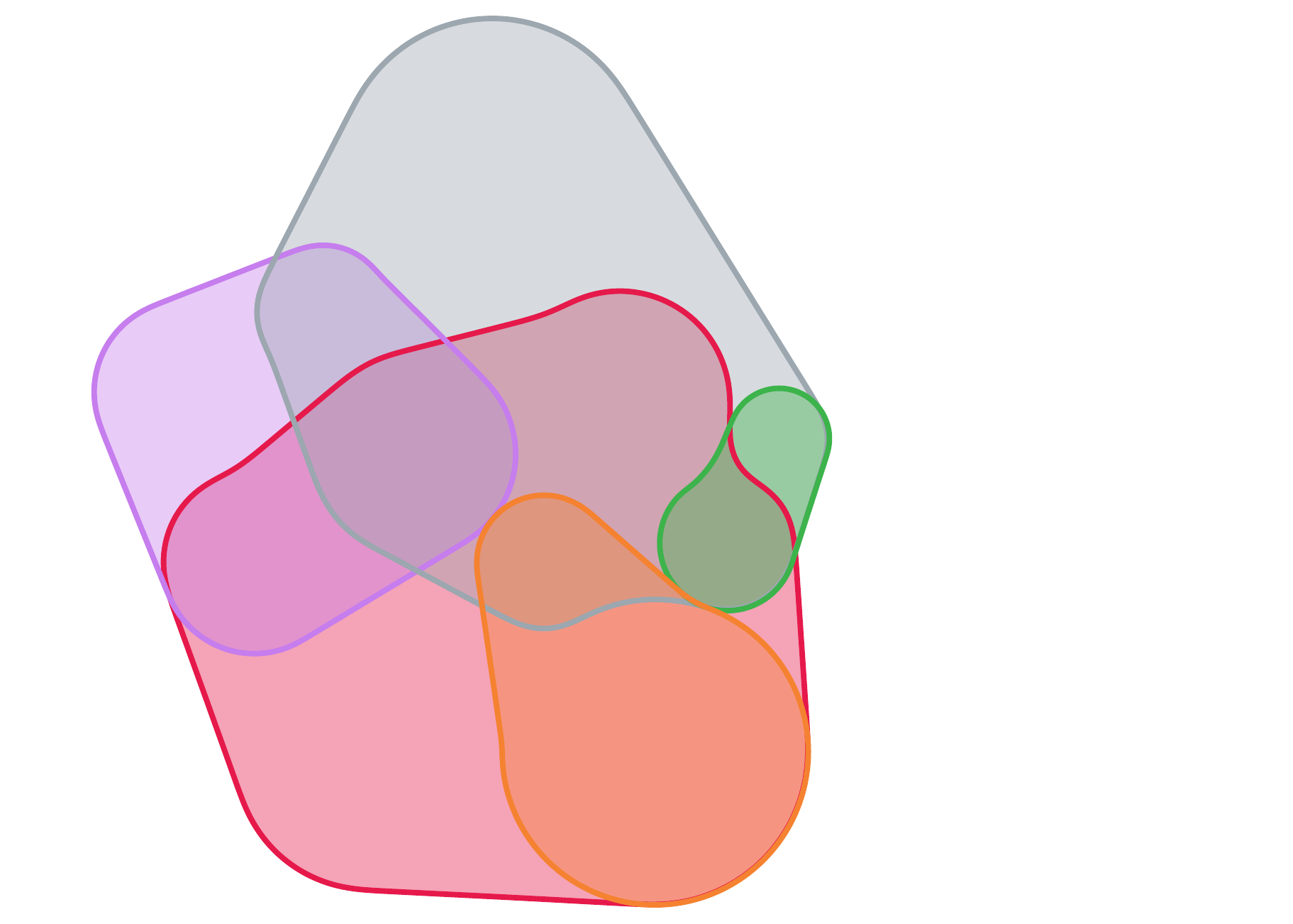
	\end{minipage}
	\caption{Venn diagrams indicating the requirement coverage (i.e., a definite result was issued) by the top-five verifiers for case study DSR (left) and ECC (right), assuming BTC results as ground truth.}
	\label{fig:venn-btc-gt}
	\vspace{-0.5cm}
\end{figure*}

Compared to Fig.~\ref{fig:quantileplot}, the scores of \kIndTool are substantially worse as it has three conflicts with BTC EmbeddedValidator.
This is due to the fact that the SV-COMP punishment scheme is bad for verifiers returning many results of which some are wrong.
It is almost as good (in terms of the scoring scheme) to not generate any result at all (and thus no ``wrong'' result).
This effect is certainly important when witness validaation is seldom, as it is the case in our setting where only one witness could be validated. 
\kIndTool produced definite verification results on many of the requirements, and consequently has a higher chance of producing a conflicting result.
With conflicts being punished heavily and non-validated answers that are deemed correct not being accounted for much, the accumulated score of a verifier returning many definitive results some of which are wrong has a high chance to score worse than a verifier returning a small number of results.
Fig.~\ref{fig:venn-btc-gt} presents updated Venn diagrams when removing all results that are in conflict with BTC.

We did a careful comparison of the verification results of all verifiers.
Our findings are summarized in Table~\ref{tab:contradictions}.
For the ECC case study, the verifiers gave contradicting answers for 18 requirements, i.e., about 24\% of all requirements. 
UltimateTaipan finds violations in 16 cases, while no verifier confirms these refutations.
There were no conflicts between the open-source verifiers for DSR.
Three conflicts were however encountered with BTC EmbeddedValidator. 
In two cases, CBMC (and \kIndTool) found a counterexample at depth two, conflicting a bounded proof of BTC EmbeddedValidator of depth 10.
As these requirements involve equality of floating-point numbers, there seems to be a subtle issue behind this.
This can be related to different intermediate floating-point precisions being used (e.g. 64 or 80 bits) and allows for multiple different, albeit correct, conflicting results.
No witnesses could be validated for any conflicting requirement, meaning that we do not have a correct measure of identifying correct answers. 
Because of the high complexity of the involved C code, we refrained from manual analysis.
While we strongly believe in the importance of witness generation and -verification, especially in industrial applications, we want to point out that in this case, the exact results are of reduced interest --- we rather want to convey the overall big picture of how well the selected open source model checkers are optimized towards real-world industrial applications.

\begin{table}[ht!]
	\caption{The contradicting results observed in  DSR and ECC, respectively.}
	\label{tab:contradictions}
	\centering
	\scriptsize
	\begin{tabular}{cp{4.4cm}p{4.4cm}r}
		\toprule
		\textbf{Case study}	& \textbf{True}				& \textbf{False}											& \textbf{Count}		\\
		\midrule
		\vspace{-0.15cm}
		\emph{DSR}			& 							& 															& 						\\
		& \kIndTool										& BTC														& 1						\\
		& BTC											& CBMC, \kIndTool											& 2						\\
		& 												& 															& $\sum = 3$			\\
		\vspace{-0.15cm}
		\emph{ECC}			& 							& 															& 						\\
		& BTC											& UltimateTaipan											& 7						\\
		& BTC, \kIndTool								& UltimateTaipan											& 4						\\
		& BTC, ESBMC, \kIndTool							& UltimateTaipan											& 4						\\
		& BTC, ESBMC									& UltimateTaipan											& 1						\\
		& ESBMC, \kIndTool								& DepthK													& 1						\\
		& ESBMC											& BTC, UltimateTaipan										& 1						\\
		& 												& 															& $\sum = 18$			\\
		\bottomrule
	\end{tabular}
	\vspace{-0.2cm}
\end{table}

\section{Encountered Issues}\label{subsec:encounteredIssues}

During the course of this work we identified issues and bugs in most of the verifiers. 
In case we were able to identify a minimal working example, we reported bugs to the developers as noted in the footnotes below. 
We give a brief description of the occurring issues.
Issues encountered with earlier versions of BTC EmbeddedValidator have been described in \cite{Berger-Ford}.

\paragraph{CBMC 5.11.} 
We encountered a bug that presented itself on the code outputted by Frama-C \cite{Cuoq-FramaC-SoftwareAnalysis}, which led CBMC to report \emph{spurious counterexamples}. 
In version 5.11, CBMC did not handle variables that are local to a switch block correctly and always assumed a non-deterministic value for them\footnote{\url{https://github.com/diffblue/cbmc/issues/3283}}. 
This bug has been fixed in subsequent releases. 
Additionally, when employing CBMC 5.9 or larger for \kIndTool we noticed \emph{a drop in performance for the inductive steps} compared to version 5.8, sometimes resulting in resource exhaustion for the 5.11 version. 
This behavior was not emerging in the base cases, i.e., it most likely corresponds to the introduced non-deterministic state spaces, although we were not able to identify a specific cause. 
Lastly, CBMC outputs \emph{witnesses that do not adhere to the format specification}\footnote{\url{https://github.com/diffblue/cbmc/issues/4418}}.

\paragraph{ESBMC 6.0.0.} 
On DSR, we observed \emph{a verifier bug on 97.1\%} of the verification tasks. 
Here, ESBMC seems to specify a faulty input for its default SMT solver, Boolector. 
Specifically, it appears to create if-else branching conditions of different sorts. 
This problem could be avoided e.g. by using Z3.

\paragraph{2LS 0.7.0.}  
We identified a simple program on which 2LS delivers \emph{false negatives}, consisting of two nested loops and a \texttt{\_\_VERIFIER\_error()} statement after the inner loop. 
2LS reports such a program as safe with its $k$-induction setting\footnote{\url{https://github.com/diffblue/2ls/issues/123}}. 
Apart from this, 2LS did not execute on any verification task. 
This seems to be due to \emph{a bug in a bit-vector map implementation}, where a size assertion fails.

\paragraph{CPAChecker 1.8.0.} C \texttt{typedef}s were not resolved correctly\footnote{\url{https://groups.google.com/forum/\#!topic/cpachecker-users/wTqHOedBOb0}}. This bug initially prevented the tool from running on the case studies completely, although \emph{it was quickly fixed by the tool developers}. 
Furthermore, we found that switch-local variables, similar to CBMC, are not represented internally at all, and thus ignored\footnote{\url{https://groups.google.com/forum/\#!topic/cpachecker-users/_bH55x_INOw}}. 
As we tried to run CPAChecker with Z3, we were deterred by a bug in the Z3-abstraction of JavaSMT\footnote{\url{https://groups.google.com/forum/\#!topic/cpachecker-users/6wv6fgwHnk4}}.

\paragraph{DepthK 3.1.} 
Due to the bug exhibited by ESBMC (see above), DepthK did not execute on most of the verification tasks. Here, it creates ESBMC instances which immediately fail until DepthK reaches the time out.

\paragraph{SMACK 1.9.3.} 
SMACK did not return a single definite answer, most likely due to the \emph{default loop bound of one}.

\paragraph{Symbiotic 6.0.3.} 
For both case studies, there are some properties for which KLEE prints that it is silently concretizing an expression to value 0 due to floating points, which leads to Symbiotic failing the verification. 
Additionally, KLEE extracted some spurious counterexamples that it could not replay. Symbiotic stops the execution thereafter.

\paragraph{UltimateAutomizer 91b1670e.} 
We observed two verification runs where UltimateAutomizer is unable to convert an assertion to an internal function representation. There are 40 ECC verification tasks leading to erroneous behavior. In 38 cases the usage of an unknown \texttt{enum} constant leads to program abortion. The remaining two instances are identical to the described bug on DSR.

\paragraph{UltimateTaipan 91b1670e.} 
On DSR and ECC, the same two conversion error instances as for UltimateAutomizer apply.

\section{Epilogue}
\label{sec:epilogue}
This paper reported on applying $12$ software model checkers to two embedded C code case studies from the automotive domain.
Although this is a rather limited set of case studies, our findings give some observations that we hope to be insightful for the software verification community. 
From the fact that the open-source verifiers cover in the best up to 20\% of all requirements ---  about 99\% of them being invariants --- makes clear that \highlight{there seems to be a serious gap between the needs of automotive code verification and open-source software model checker capabilities}.
The specific characteristics of the two case studies (many floating-points, pointer dereferencing, bitwise operations etc.) are certainly a decisive factor in this respect.
Additionally, the structure of an infinite outer loop (forever processing inputs) with nested finite loops seems to require an tailored k-induction to properly capture behavior, which we believe explains part of the success of \kIndTool and BTC.
While both tools are heavily tailored towards special use-cases and are unsuitable for more general programs, we firmly believe these optimizations are worth pursuing and integrating into mainstream open-source verifiers.
Admittedly, the fact that our benchmarks are not publicly available is a weak point. 
More studies like the one in this paper are needed. 
To that end, \highlight{the software model checking community and industrial partners covering various application domains should take up an orchestrated effort to set up a substantial set of industrial benchmarks}.
The only way to meet the needs in industry is to be able to apply software model checkers on real industrial software of different domains.
Finally, the results of our study (particularly, the score of \kIndTool relative to BTC) suggest \highlight{to revisit the scoring scheme of verification competitions such as SV-COMP}.
In particular, the punishment of wrong verification results is too severe; it is currently measured in absolute terms (the number of wrong answers), whereas a relative judgment (what is the percentage of wrong answers that a verifier obtained) seems to be more fair.

\paragraph*{Acknowledgments.}\label{sec:Acknowledgments}
\ifanon
Removed for double-blind review.
\else
We thank BTC Embedded Systems AG, in particular Tino Teige and Markus Gros, for their support and helpful advice. 
We are grateful to Md Tawhid Bin Waez and Thomas Rambow (both from Ford Motor Company) for their support on the case studies in an earlier phase and for fruitful discussions on formal verification and Simulink. 
We thank Dirk Beyer for very useful feedback on an earlier version of the paper.
\fi

\ifappendix

\newpage
\begin{subappendices}
\renewcommand{\thesection}{}%
\section{}
\label{appendix}
Although powerful benchmarking tools, e.g. BenchExec\footnote{\url{https://github.com/sosy-lab/benchexec}}, exist, our infrastructure forced us to use a custom benchmark script\ifanon\footnote{\url{https://github.com/icst-2020-submission/benchmarking-verifiers-for-automotive}}\else\footnote{\url{https://github.com/lu-w/benchmarking-verifiers-for-automotive}}\fi. Despite not being as powerful and versatile as BenchExec, it still provides the following basic functionality: Executing multiple verification runs in a parallel fashion, assigning exclusive CPU cores to each run via \texttt{taskset}, restricting memory and CPU-time of each run while controlling for possible parallelism of the verifiers by using \texttt{timeout}\footnote{\url{https://github.com/pshved/timeout}}, taking measurements of time and memory from the output of \texttt{timeout}, interpreting verifier output, and storing the data in a CSV file. 

Our setup does not control for hyper-threading on the employed Intel CPUs, as this feature was disabled on our infrastructure. Possible influences between processes are mitigated by using distinct cores for each verification run. The following tool configurations and parameters were used:

\texttt{2ls --32 --heap-interval --k-induction}\\
\texttt{cbmc --stop-on-fail --32 --unwindset main.0:}$k$\\
\texttt{cpa.sh -stats -32 -timelimit 999999s -heap 18G -spec reach.prp\\\hspace*{0.8cm} -svcomp19}\\
\texttt{depthk.py --force-check-base-case -a 32 -t 7200 --solver boolector\\\hspace*{0.8cm}--prp reach.prp --extra-option-esbmc='--floatbv\\\hspace*{0.8cm}--context-bound 2'}\\
\texttt{esbmc --no-div-by-zero-check --force-malloc-success\\\hspace*{0.8cm}--state-hashing --no-align-check --k-step 2 --context-bound 2\\\hspace*{0.8cm}--32 --floatbv --k-induction --no-pointer-check\\\hspace*{0.8cm}--no-bounds-check --interval-analysis --unlimited-k-steps}\\
\texttt{kinduction.py --config ../config/cbmc.yaml --variable-moving\\\hspace*{0.8cm}--ignore-functions-for-variable-moving updateVariables}\\
\texttt{pesco.sh -stats -32 -heap 18G -spec reach.prp -svcomp19-pesco}\\
\texttt{smack --clang-options=-m32 --float -v --bit-precise --time-limit\\\hspace*{0.8cm} 7200}\\
\texttt{symbiotic --32}\\
\texttt{Ultimate.py --data data --config config --spec config/svcomp-\\\hspace*{0.8cm}Reach-32bit-Automizer\_Bitvector.epf --architecture 32bit}\\
\texttt{Ultimate.py --data data --config config --spec config/svcomp-\\\hspace*{0.8cm}Reach-32bit-Kojak\_Bitvector.epf --architecture 32bit}\\
\texttt{Ultimate.py --data data --config /config --spec config/svcomp-\\\hspace*{0.8cm}Reach-32bit-Taipan\_Bitvector.epf --architecture 32bit}\\
The file \texttt{reach.prp} contains \texttt{CHECK(init(main()), LTL(G ! call(\_\_VERIFIER\\\_error())))}.\\
For CPAChecker and PeSCo, the \texttt{svcomp19} and \texttt{svcomp19-pesco} configurations were scaled for our specified run times, meaning that each configured time limit was multiplied by eight. For CBMC, $k$ was replaced by increasing values.

\end{subappendices}
\fi
\newpage
\bibliographystyle{splncs}
\bibliography{references}

\end{document}